\documentclass{aa}  
\usepackage{graphicx}
\usepackage{txfonts}
\newcommand{\ket}[1]{|#1\rangle}

\newcommand{\bra}[1]{\langle #1|}

\begin{document}
    \title{Fundamental Limits to Phase and Amplitude Estimation in the High-Strehl Regime}

    \author{Jacob Trzaska\inst{1}\fnmsep\thanks{Corresponding author: \email{jtrzaska@arizona.edu}}
        \and
        Amit Ashok\inst{1, 2}
    }

    \institute{
        Wyant College of Optical Sciences, University of Arizona, Tucson, Arizona, 85721, USA\and
        Department of Electrical and Computer Engineering, University of Arizona, Tucson, Arizona, 85721, USA\\
    }

\date{\today}

\abstract{}{}{}{}{}
\abstract
{Ground-based telescopes are susceptible to seeing, an atmospheric phenomenon that reduces the resolving power of large observatories to that of a home telescope. Compensating these effects is therefore critical to realizing the potential of upcoming extremely large telescopes, a challenging task that requires precise wavefront control. Ultimately, this precision is limited by one's wavefront sensor (WFS) and its capacity to accurately encode phase and amplitude aberrations.}
{Our attention is on photon noise-limited wavefront sensing in the high-Strehl regime. In particular, we seek fundamental limits to phase and amplitude estimation in addition to a WFS that saturate these bounds.}
{Information theory is employed for deriving minimum-achievable residual errors, as stipulated by a metric called the Holevo Cramér-Rao bound. Holevo's bound is closely related to another metric called the quantum Cramér-Rao bound, which has already been applied to phase estimation on nearly-corrected wavefronts.}
{We present a WFS that can perfectly extract and phase shift a telescope's piston mode. We show how this phase can be used to tune the apparatus' sensitivity to phase and amplitude, and provide a closed-form expression for the optimal phase shift. For circular apertures, this implementation saturates the fundamental limits, but it can be easily modified to work with arbitrary pupils. Moreover, our proposal uses optics that are manufactureable today and is readily achromatized with geometric phase shifters.}
{}
\keywords{
    Instrumentation: adaptive optics --
    Instrumentation: detectors
}

\maketitle
%

\section{Introduction}
Ground-based telescopes are susceptible to seeing, an atmosphere-induced blur that limits angular resolution to a few arcseconds, regardless of the observatory's size and optical quality. Seeing poses a significant observational challenge, especially with regard to applications such as high-contrast imaging, where the search for extra-solar planets demands probing potential star-planet separations well below one arcsecond \citep{guyon2018extreme}. These requirements have driven significant research into extreme adaptive optics (exAO), closed-loop control systems that monitor and correct atmospheric phase and amplitude aberrations across thousands of modes in sub-millisecond timescales. At the core of exAO is a wavefront sensor (WFS), the device responsible for encoding phase and amplitude errors as interpretable intensities. Ideally, WFS measurements report an unambiguous signal exactly specifying any atmospheric distortions, ultimately enabling perfect correction.  

However, photon shot noise introduces an intrinsic uncertainty to the measurement, fundamentally limiting AO performance by checking WFS sensitivity. Indeed, for a given WFS $\Pi$ and $N$ collected photons, one can define a metric called the Fisher information matrix (FIM) $\mathcal{F}(\Pi)$ \citep{demkowicz-dobrzanski_multi-parameter_2020, haffert2023}, which lower bounds the residual mean-squared wavefront error $\text{Tr}(\Sigma)$ via the Cramér-Rao bound (CRB):
\begin{equation}
    \label{eq:ccrb}
\text{Tr}(\Sigma) \geq \frac{1}{N}\text{Tr}\left[\mathcal{F}^{-1}(\Pi)\right].
\end{equation} Consequently, a finite residual wavefront always persists through observations, limiting achievable angular resolution in addition to post-coronagraph contrast. 

If we model the wavefront sensing problem quantum-mechanically then it is possible to optimize the FIM over \textit{all} conceivable WFS designs. In this manner we can construct an absolute ruler with which to benchmark different WFS and wavefront reconstruction algorithms. Moreover, such a result can be used to construct realistic expectations for ground-based high-contrast imaging. Recent work \citep{haffert2023} has already identified a quantum bound for estimating the expansion coefficients of a wavefront as expressed in an orthonormal basis. See also \citet{paterson_towards_2008}, where the author proposed phase shifting the unperturbed field by $\pi/2$ to saturate the limit to an imaging WFS. \citet{Trzaska:24} showed that spatial mode sorting can also saturate this limit. Soon after, \citet{chambouleyron_coronagraph-based_2024} presented an alternative and concrete approach based on Fourier filtering with a bi-vortex phase mask. Numerical simulation showed that the latter has near-optimal sensitivity at high spatial frequencies, though a nontrivial gap exists at low frequencies. 

An optimal WFS thus remains absent, as well as a quantum optics-informed study detailing fundamental limits to amplitude estimation. The latter's importance stems from an ideal phase sensor being completely blind to amplitude errors and vice-versa (see Section \ref{sec:joint-estimation}). Here, we address these outstanding challenges through a rigorous quantum information analysis, presenting both new bounds and the piston-adapted WFS (PAWS), a new WFS that saturates the fundamental limits to simultaneous phase and amplitude estimation. Throughout this work we have assumed (i) photon noise-limited measurements and (ii) that we are operating in the high-Strehl regime. (Electronic noises as accrued by dark current and detector readout are ignored as they can be driven down with proper cooling and gain.) As such, we expect PAWS to find application in second-stage AO systems, further driving down wavefront errors after the loop has been closed.

\section{Fundamental Limits}
\label{sec:opt-wfs}
We consider a telescope having a binary pupil mask $\mathcal{P}$ that is staring at a distant on-axis point source (guide star). This point-like assumption implies that only a single optical mode $\psi(\mathbf{x})$ is excited within the pupil at any given time. Hence, the field is spatially coherent with itself and can be deterministically expanded as \begin{equation}
    \label{eq:abstract-mode}
    \psi(\mathbf{x}) = e^{l(\mathbf{x}) + i\phi(\mathbf{x})},
\end{equation} where $l$ and $\phi$ are the field's log-amplitude and phase, respectively. In the high-Strehl regime both these functions are closely approximated as piston modes, with log-amplitude further constrained by the normalization\begin{equation}
    \label{eq:normalization}
    \int_{\mathcal{P}}e^{2l(\mathbf{x})}d\mathbf{x} = 1.
\end{equation}Equation \eqref{eq:normalization} lets us interpret $|\psi(\mathbf{x})|^2$ as a probability density for measuring a photon at the position $\mathbf{x}$.

To facilitate wavefront estimation we will expand both phase and log-amplitude using linearly-independent functions. In particular, let us employ expansions in orthogonal functions, viz., \begin{align}
    \label{eq:ampl-expansion}
    l(\mathbf{x}) &= \sum_{n=0}^{\infty}l_{n}\chi_{n}(\mathbf{x})\\
    \label{eq:phase-expansion}
    \phi(\mathbf{x}) &= \sum_{n=0}^{\infty}\phi_{n}\chi_{n}(\mathbf{x}),
\end{align} where $\{\chi_{n}\}_{n=0}^{\infty}$ are constructed such that 
\begin{equation}
    \label{eq:ip-def}
    \begin{aligned}
    \left\langle\chi_{k}, \chi_{l}\right\rangle_{e^{2l}} &= \int_{\mathcal{P}}e^{2l(\mathbf{x})}\chi_{k}(\mathbf{x})\chi_{l}(\mathbf{x})d\mathbf{x}\\
    &= \delta_{kl},
    \end{aligned}
\end{equation} and $\chi_0$ is constant over the aperture. Phase and (log-) amplitude estimation now amounts to estimating the numbers $\{l_n\}_{n=1}^{\infty}$ and $\{\alpha_n\}_{n=1}^{\infty}$, where $n=0$ coefficients are excluded since $\phi_0$ is just a global piston and $l_0$ is determined by normalization. Because are we are operating in the high-Strehl limit, we are really estimating field perturbations about a perfect wavefront, further implying \begin{equation}
    |l_n|, |\phi_n| << 0\,\,\,\,\,(n\geq1).
\end{equation}

\subsection{Quantum Cramér-Rao Bounds}
\label{sec:qcrb}
Quantum information theory is concerned finding a WFS and estimator pair that simultaneously minimizes $\text{Tr}(\mathcal{C}\Sigma)$ \emph{and} $\text{Tr}[\mathcal{C}\mathcal{F}^{-1}(\Pi)]$. The non-negative matrix $\mathcal{C}$ appearing here weights the importance of different parameters. For example, if we are estimating $[\phi_1,\dots,\phi_K, l_1,\dots,l_K]^T$, then \begin{align}
    \mathcal{C} &= \begin{bmatrix}
        I_{K\times K} & 0\\
        0 & 0
    \end{bmatrix},\\
    \mathcal{C} &= \begin{bmatrix}
        0 & 0 \\
        0 & I_{K\times K}
    \end{bmatrix},
\end{align}will produce a bound considering only phase or amplitude, respectively. (The symbol $I_{K\times K}$ is a K by K identity matrix.) In this work we will consider weight matrices of the form \begin{equation}
    \mathcal{C} = \begin{bmatrix}
        I_{K\times K} \cos^2\Theta & 0\\
        0 & I_{K\times K} \sin^2\Theta
    \end{bmatrix}
\end{equation} and seek the corresponding limit.

Optimizing the estimator covariance and WFS sensitivity has traditionally relied on arguments related on somewhat heuristic sensitivity definitions \citep{guyon_limits_2005, chambouleyron_variation_2021}. In contrast, quantum information theory gives a rigorous and systematic prescription for optimizing both. Its principal tool is the quantum Fisher information (QFI; $\mathcal{Q}$) \citep{helstrom_minimum_1968}, which lower bounds the right side of \eqref{eq:ccrb} via the quantum Cramér-Rao bound (QCRB) \begin{equation}
    \label{eq:qcrb}
    \text{Tr}(\mathcal{C}\Sigma) \geq \frac{1}{N}\text{Tr}\left[\mathcal{C}\mathcal{F}^{-1}(\Pi)\right] \geq \frac{1}{N}\text{Tr}\left[\mathcal{C}\mathcal{Q}^{-1}\right].
\end{equation} For single-conjugate wavefront sensing, the QFIM can be written component-wise as \citep{liu_quantum_2020} \begin{equation}
    \label{eq:pure-state-qfi}
    \mathcal{Q}_{nm} = 4\text{Re}\left\{\langle\partial_n\psi,\partial_m\psi\rangle_1 - \langle\partial_n\psi,\psi\rangle_1\langle\psi,\partial_m\psi\rangle_1\right\},
\end{equation} where the $n$ and $m$ indices may refer to either a phase or amplitude coefficient. Substituting \eqref{eq:abstract-mode} into \eqref{eq:pure-state-qfi} immediately yields \begin{equation}
    \label{eq:total-qfim}
    \mathcal{Q} = 4I,
\end{equation}where $I$ here is an infinite-dimensional identity matrix. When considering phase-only estimation, equation \eqref{eq:total-qfim} reproduces the results in \citet{haffert2023}. When considering (log-) amplitude-only estimation, we arrive at the new result that log-amplitudes are estimable to a precision of 0.5 rad. With $N$ photons the precision improves by $\sqrt{N}$.

Intuitively, QFI measures the overlap between classical variations in intensity and our aberration modes. Because a good wavefront sensor linearly translates phase variation into intensity variation, sensitivity is maximized when fringe contrast is maximized. This fringe contrast has previously been shown equal to 2 \citep{chambouleyron_coronagraph-based_2024}, which is related by the QFIM through a square root relation \citep{haffert2023}. This intuition also explains the second term in \eqref{eq:pure-state-qfi}, as the fraction of light overlapping with piston produces zero signal.

\subsection{Holevo Cramér-Rao Bound}
\label{sec:joint-estimation}
While QFI satisfies our intuition on what a good sensitivity metric should be, it can be overly-optimistic. Notice that \eqref{eq:pure-state-qfi} is defined as the real part of an otherwise complex matrix. When this auxiliary matrix is real, the QFI (and thus QCRB) is saturable and there exists a \emph{quantum-optimal} WFS that reaches the fundamental limit \citep{pezze_optimal_2017}. However, when this auxiliary matrix is complex, the second inequality in \eqref{eq:qcrb} is strict. In essence, we have discarded information contained in the imaginary component. This is analogous to self-coherence measurements in a double pinhole interferometer. There measuring a single set of fringes only yields the coherence function's real part. Completely characterizing the field's statistics requires at least one additional measurement to pin down imaginary terms. The upshot is that additional processing is required, and this fundamentally eats away at available sensitivity. 

For single-conjugate wavefront sensing this is unfortunately the case. Indeed, terms in equation \eqref{eq:pure-state-qfi} that are indexed by both phase and amplitude coefficients are always complex, viz., \begin{align}
    \text{Im}\langle\partial_{\phi_n}\psi,\partial_{l_m}\psi\rangle &= -\text{Im}\left[i\int_{\mathcal{P}}e^{2l(\mathbf{x})}\chi_{n}(\mathbf{x}) \chi_{m}(\mathbf{x})d\mathbf{x}\right]\\
    &=-\delta_{nm}. \label{impossible:eq}
\end{align} That the joint QCRB is unachievable can also be seen from its very definition. Let $\Psi$ be the post-WFS optical field just prior to detection, i.e., $\Psi\equiv\Pi(\psi)$. We assume that $\Pi$ is unitary. The amplitude FIM is then given by\begin{equation}
    \label{eq:fa-final}
    \mathcal{F}_A\left(\Pi\right) = 4\int_\mathcal{P}\frac{\text{Re}\left[\Psi^*(\mathbf{x})\nabla \Psi(\mathbf{x})\right]\text{Re}\left[\Psi^*(\mathbf{x}) \nabla^T \Psi(\mathbf{x})\right]}{|T(\psi)_k|^2}d\mathbf{x},
\end{equation}where $\nabla=[\partial_{l_1}, \partial_{l_2}, \dots, \partial_{l_M}]^T$, and the phase FIM by \begin{align}
    \mathcal{F}_P\left(\Pi\right) &= 4\int_\mathcal{P}\frac{\text{Re}\left[i\Psi^*(\mathbf{x})\nabla \Psi(\mathbf{x})\right]\text{Re}\left[i\Psi^*(\mathbf{x}) \nabla^T \Psi(\mathbf{x})\right]}{|\Psi(\mathbf{x})|^2}d\mathbf{x}\\
    \label{eq:fp-final}
    &= 4\int_\mathcal{P}\frac{\text{Im}\left[\Psi^*\nabla \Psi(\mathbf{x})\right] \text{Im}\left[\Psi^*(\mathbf{x})\nabla^T \Psi(\mathbf{x})\right]}{|\Psi(\mathbf{x})|^2}d\mathbf{x}
\end{align} Equations \eqref{eq:fa-final} and \eqref{eq:fp-final} together imply \begin{align}
    \mathcal{F}_A + \mathcal{F}_P &= 4\int_\mathcal{P}\frac{\left[\Psi^*(\mathbf{x})\nabla \Psi(\mathbf{x})\right]\left[\Psi(\mathbf{x})\nabla^T\Psi^*(\mathbf{x})\right]}{|\Psi(\mathbf{x})|^2}d\mathbf{x}\\
&=4\int_\mathcal{P}\nabla \Psi(\mathbf{x})\nabla^T \Psi^*(\mathbf{x})d\mathbf{x}\\
    \label{eq:fundamental-tradeoff}
    &= 4I_{K\times K}
\end{align} or, more colloquially, that there exists a fixed sensitivity budget that must be purposefully allocated between the phase and amplitude estimation tasks. This allotment is exactly \emph{one} QFIM, rather than two, as might be naively expected, and stems from the incommensurate manner in which one saturates each bound. Equation \eqref{eq:fundamental-tradeoff} not only proves that the QCRB is unduly auspicious, but also that QCRB-saturating phase sensors are wholly insensitive to amplitude and vice-versa. (An identical result also holds for spatial mode sorting measurements.)

To derive a truly fundamental limit we must instead turn to more sophisticated quantum information metrics. In particular, we will find Holevo's Cramér-Rao bound (HCRB)\begin{equation}
    \label{eq:holevo-definition}
    \text{Tr}(\mathcal{C}\Sigma) \geq \frac{1}{N} \mathcal{H},
\end{equation} suitable for wavefront sensing, as it always saturable for the weak starlight collected by telescopes. (In quantum mechanical parlance, starlight is approximately a ``pure state." \citep{TsangQSR}) The constant $\mathcal{H}$ specifies our per-photon sensitivity, its calculation is a difficult task that requires somewhat mature understanding of quantum information theory. For this reason, we have deferred a detailed mathematical exposition to Appendix \ref{app:sec:HCRB}. Instead, we shall simply quote the fundamental inequality as \begin{equation}
    \label{eq:holevo-bound}
    \text{Tr}(\mathcal{C}\Sigma) \geq \frac{K}{4N}\left[1 + \sin(2\Theta)\right],
\end{equation}where $K$ is the number of modes being estimated. 

It is clear that \eqref{eq:total-qfim} produces a lower bound that is $K\sin(2\Theta)/N$ \emph{greater} than a CRB derived from \eqref{eq:total-qfim}. Whenever $\Theta=0$ or $\pi/2$ we immediately recover the scalar bound corresponding to the phase-only and amplitude-only estimation, respectively. However, the incompatibility between phase and amplitude is apparent for an equal weighting ($\Theta=\pi/4$), which produces a lower bound twice that naively expected from \eqref{eq:total-qfim}. Indeed, for $0 < \Theta < \pi/2$, Holevo's bound is always larger than that derived from the quantum Fisher information. This stems from our having two produce two overlapping interference patterns in the pupil plane. One of these fringe set encodes the pupil field's amplitude perturbations and the other its phase variation.

\subsubsection{Strehl Ratio\label{sec:Strehl-ratio}}
\begin{figure}
    \centering
    \includegraphics[width=\linewidth]{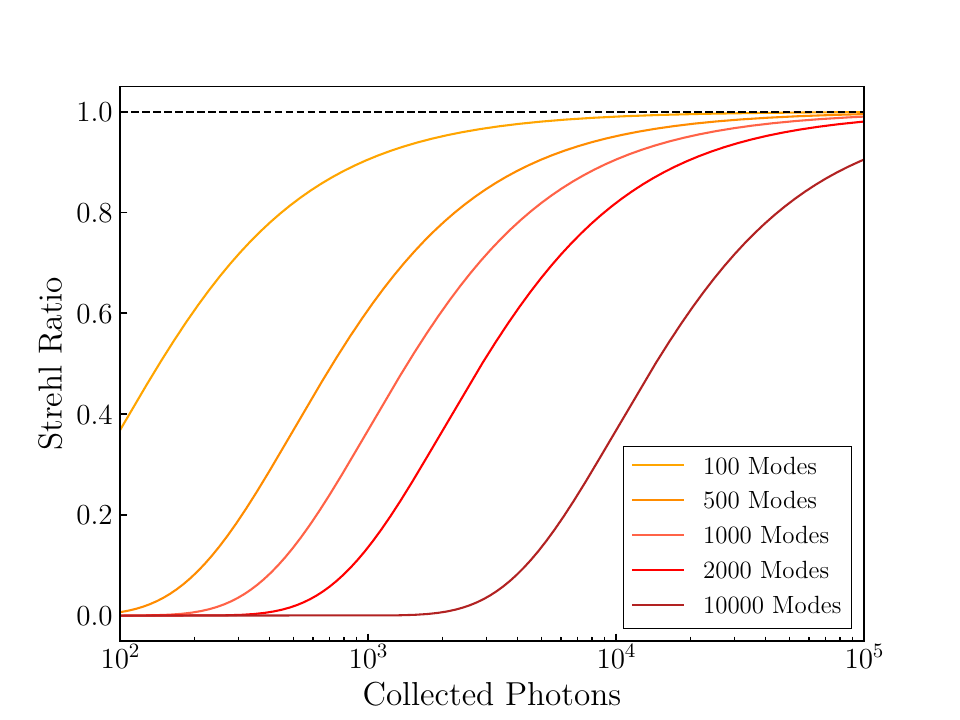}
    \caption{Expected long-exposure Strehl ratio against the number of photons collected and modes corrected. Plots are shown for an optimal WFS measuring phase and amplitude with equal weight. We see that adaptive optics has little utility when collecting fewer than four photons for every ten modes. Conversely, Strehl ratios of around of 80\% are achievable with about 4.5 photons per mode.}
    \label{fig:correction-limits}
\end{figure}

Equation \eqref{eq:holevo-bound} can also be understood in terms of the long-exposure Strehl ratio \citep{ross_limitations_2009}\begin{equation}
    S = \exp\left(-\sigma_l^2 - \sigma _{\phi}^2\right),
\end{equation} where we have invoked the extended Maréchal approximation. (If AO is operating at the quantum limit then its residual error is Gaussian and this approximation is exact.) Unlike mean-squared error, Strehl ratio can be readily measured from a focal plane images and used to monitor image quality. In this sense it is more operationally relevant. 

Figure \ref{fig:correction-limits} predicts Strehl ratios that can be expected from quantum-limited estimation (and ideal correction) against the number of collected photons. Diffraction-limited performance ($S\geq 0.8$) is achieved with approximately 4.5 photons per mode. We contrast this with the single photon per mode criterion derived by  \citet{paterson_towards_2008} for phase-only sensing. Achieving even a moderate level of AO correction, say $S=10\%$, will require more than four photons for every ten modes. As a concrete example, reaching these Strehls with a 3 kHz AO loop that is correcting for 20,000 modes on a 25.4 m telescope would require an 8.69 magnitude guide star (calculated using values from \citet{LGS-Flux}). This is a demanding quantity, especially in a coronagraphy context where AO systems are limited to using a potential exoplanet's host star for wavefront sensing. 

\section{Piston-Adapted Wavefront Sensor}
\begin{figure*}
    \centering
    \includegraphics[width=\linewidth]{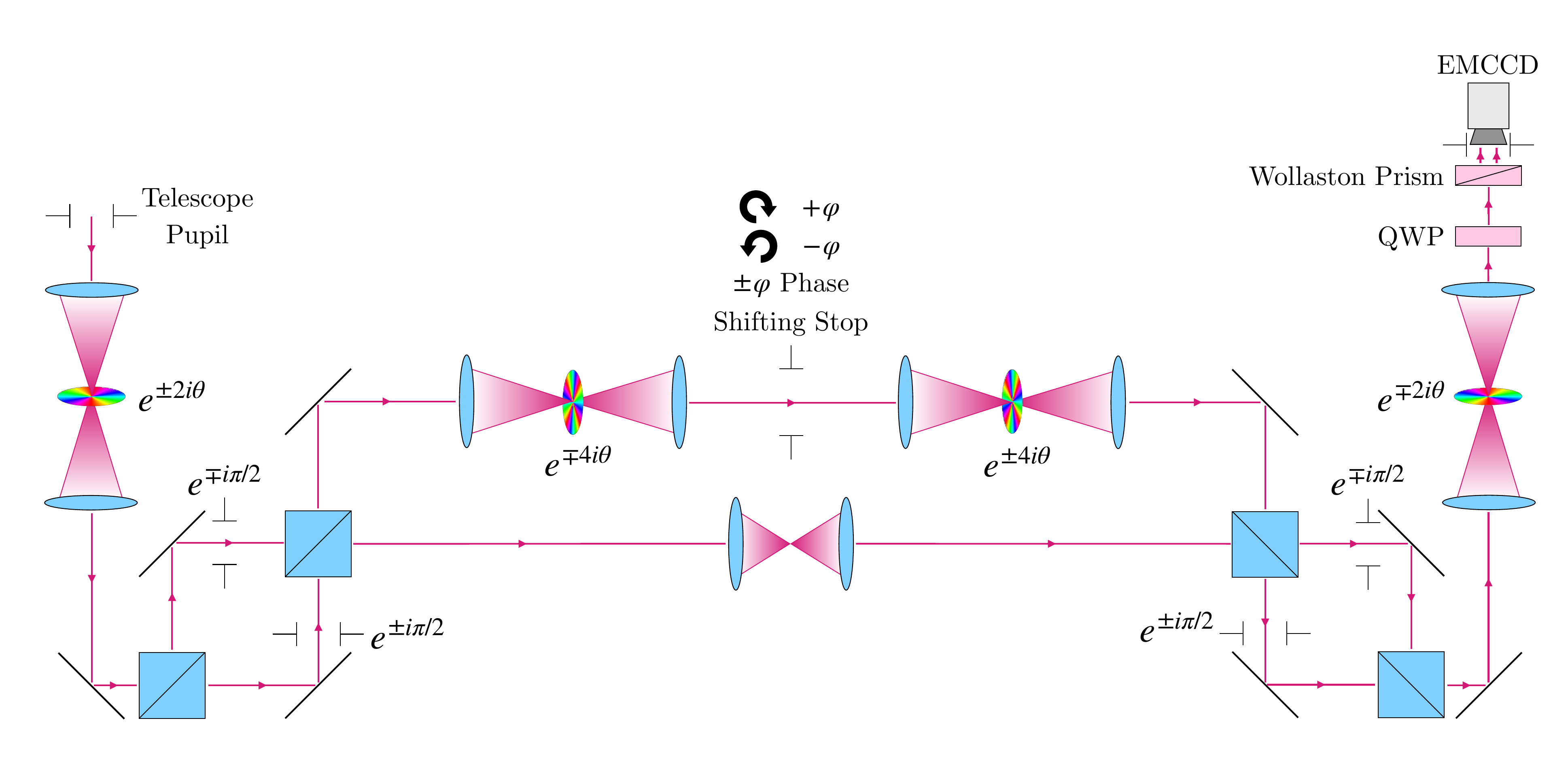}
    \caption{The piston-adapted WFS (PAWS). Our construction spatially sorts the complex Zernike polynomials in two stages, first using a bi-vortex phase filter and Mach-Zhender interferometer (MZI) to separate $m=\pm n$ modes from the remainder, and then separating piston from high-order $m=\pm n$ polynomials using a quad-vortex mask. We envision an implementation using vector vortex masks, which introduce conjugate phases to left- and right-handed circular polarizations states (hence the $\pm$). Likewise, both apertures inside the MZIs use geometric phase to shift ejected modes. Sensitivity to phase and amplitude is then controlled using another half-wave plate with a hole cut to the aperture size. Rotating this waveplate changes the geometric phase $\varphi$, increasing sensitivity to phase aberrations as $\varphi\rightarrow\pi/2$ or amplitude as $\varphi\rightarrow0$. At $\varphi=\pi/4$ we allocate equal sensitivity between the two aberrations, minimizing the total residual variance. Like the vZWFS, our WFS works on the two orthogonal circular polarizations, imparting conjugate phases to each. Optics right of the figure's center serve to reconstitute the field at the pupil plane and produce two images, one for each circular polarization, via a Wollaston prism. A polarizing cube beamsplitter could also be used. Note that this figure is adapted from \citet{trzaska2025zernikemodesortingvortex}.}
    \label{fig:holevo-optimal}
\end{figure*}

It remains to develop hardware saturating inequality \eqref{eq:holevo-bound}. To this end we introduce the piston-adapted WFS (PAWS), to our knowledge the most sensitive WFS ever proposed (Figure \ref{fig:holevo-optimal}). Our WFS takes inspiration from the bi-vortex WFS introduced by \citet{chambouleyron_coronagraph-based_2024} and exploits the action of vortex phase filters on complex Zernike polynomials as articulated in \citet{trzaska2025zernikemodesortingvortex}.

Our implementation should be compared to the vector Zernike WFS (vZWFS), which uses liquid crystal mask to impart a $\pm\pi/2$ to the PSF core on two orthogonal circular polarization states. This same technology can be used to implement the vortex phase mask \citep{mawet_optical_2009, mawet_vector_2010}. Also like the vZWFS, we use a Wollaston prism to produce two separate images of the pupil, with each image corresponding to a different circular polarization. For small perturbations to the incident field, these intensity measurements are linearly related to the phase and amplitude aberrations via \begin{subequations}
\label{eq:paws-intensities}
\begin{align}
    I_\circlearrowright(\mathbf{x}) &\propto 1 + l(\mathbf{x})\cos\varphi + \phi(\mathbf{x})\sin\varphi + \mathcal{O}\left(l^2\right)+\mathcal{O}\left(\phi^2\right)\\
    I_\circlearrowleft(\mathbf{x}) &\propto 1 + l(\mathbf{x})\cos\varphi - \phi(\mathbf{x})\sin\varphi + \mathcal{O}\left(l^2\right) + \mathcal{O}\left(\phi^2\right).
\end{align}\end{subequations}Hence, the output intensities are related to phase and amplitude at each point via a beamsplitter-type relation. In very-high Strehl scenarios we can therefore expect a linear reconstruction algorithm to offer good performance.

In principle the entire optical train is easily achromatized. Lenses shown in Figure \ref{fig:holevo-optimal} can be replaced with off-axis parabolic mirrors. Mach-Zehnder interferometers (MZIs) are inherently achromatic when used with pellicle beamsplitters. For the phase shifters we employ half-wave plates with rotated fast axes. Achromatic waveplates are available from most large optics vendors; when their fast axes are rotated by $\gamma$, we induce broadband phase shifts $\exp(\pm2i\gamma)$ to right- and left-handed circularly polarized photons. With these design choice, both the vortex filters and interferometers are rendered broadband.

PAWS' optimality for $\Theta=0$ follows from results in \citet{chambouleyron_coronagraph-based_2024}. Optimality for $\Theta=\pi/2$ is shown in Appendix \ref{app:sec:amplitude}. In this case PAWS is just a relay and uses all the light to form a pupil image. Optimality for $0<\Theta<\pi/2$ is proven in Appendix \ref{app:sec:HCRB-PAWS}. In general, one can show that choosing \begin{equation}
    \varphi = \tan^{-1}\sqrt{\cot\Theta}
\end{equation} gives PAWS a sensitivity equaling the fundamental limit. (Experimentally, this means rotating the half-wave plate by $\varphi/2$.)

That said, PAWS as presented is only optimal when used with single-frequency light. A broadband quantum-limited sensor must spectrally disperse the light and measure filtered pupil images at each wavelength. This additional complication affords extra performance, but the magnitude of these benefits will depend on the guide star's power spectral density as well as the atmosphere's dispersion profile. Advantages will also be marginal for small bandwidths. 

PAWS is also only optimal for circular apertures. This hurdle can be circumvented by pre-processing the incident field with a single-mode converter prior to entering PAWS, and then placing the inverse device afterwards \citep{trzaska2025zernikemodesortingvortex}. In particular, we want to convert the aperture's native piston mode into Zernike piston, which we have shown can be precisely controlled. At least for quasimonochromatic light, such as a transformation can be accomplished with high fidelity using a few spatially separated phase planes \citep{oliker_beam_2018, raeker_lossless_2021}. This optical pre-filter thus extends PAWS' quantum-optimality to arbitrary apertures, including next-generation observatories such as the Giant Magellan and European Extremely Large telescopes.

\subsection{Comparative Sensitivity Analysis}
\label{sec:measurements}
\begin{figure}
    \centering
    \includegraphics[width=\linewidth]{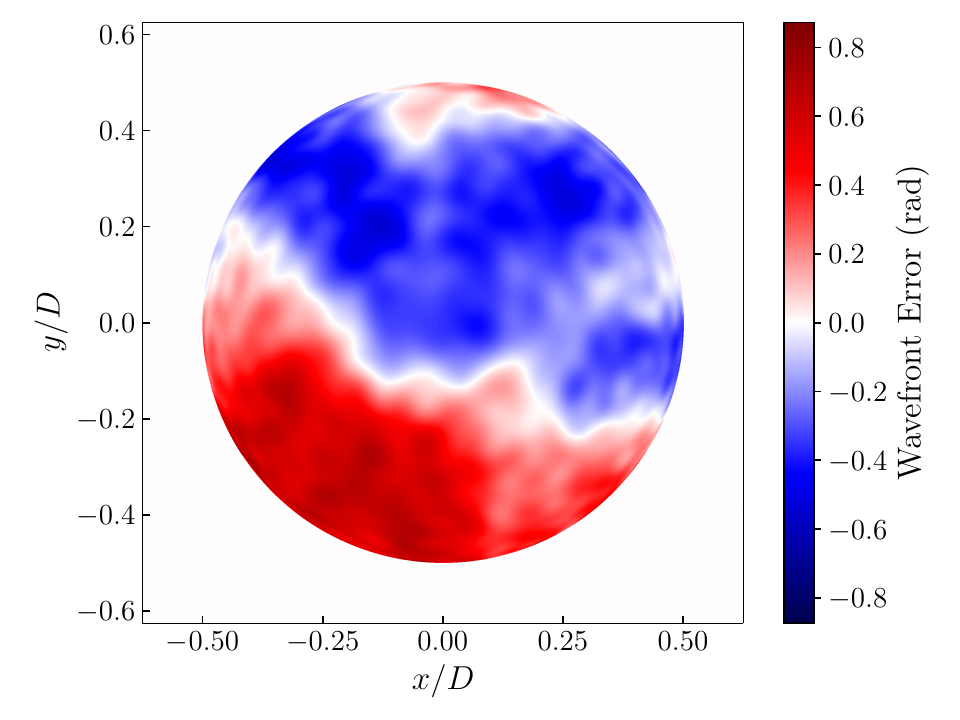}
    \caption{A circular-symmetric imaging system of diameter $D$ with atmospheric phase overlaid.}
    \label{fig:single-aperture-geometry}
\end{figure}

Before closing, we will numerically compare the sensitivity of various wavefront sensors against a PAWS optimized for (i) phase, (ii) amplitude, and (iii) equally-weighted phase and amplitude sensing. For this analysis we have chosen to evaluate comparative sensitivities for an unmodulated, four-sided pyramid WFS (PyWFS) and a vZWFS with $2\lambda/D$ dot diameter. PyWFS simulations possessed maximum surface ratio \citep{fauvarque_general_2017}. Our aperture is the circular pupil shown in Figure \ref{fig:single-aperture-geometry}.

\begin{figure}
    \centering
    \includegraphics[width=\linewidth]{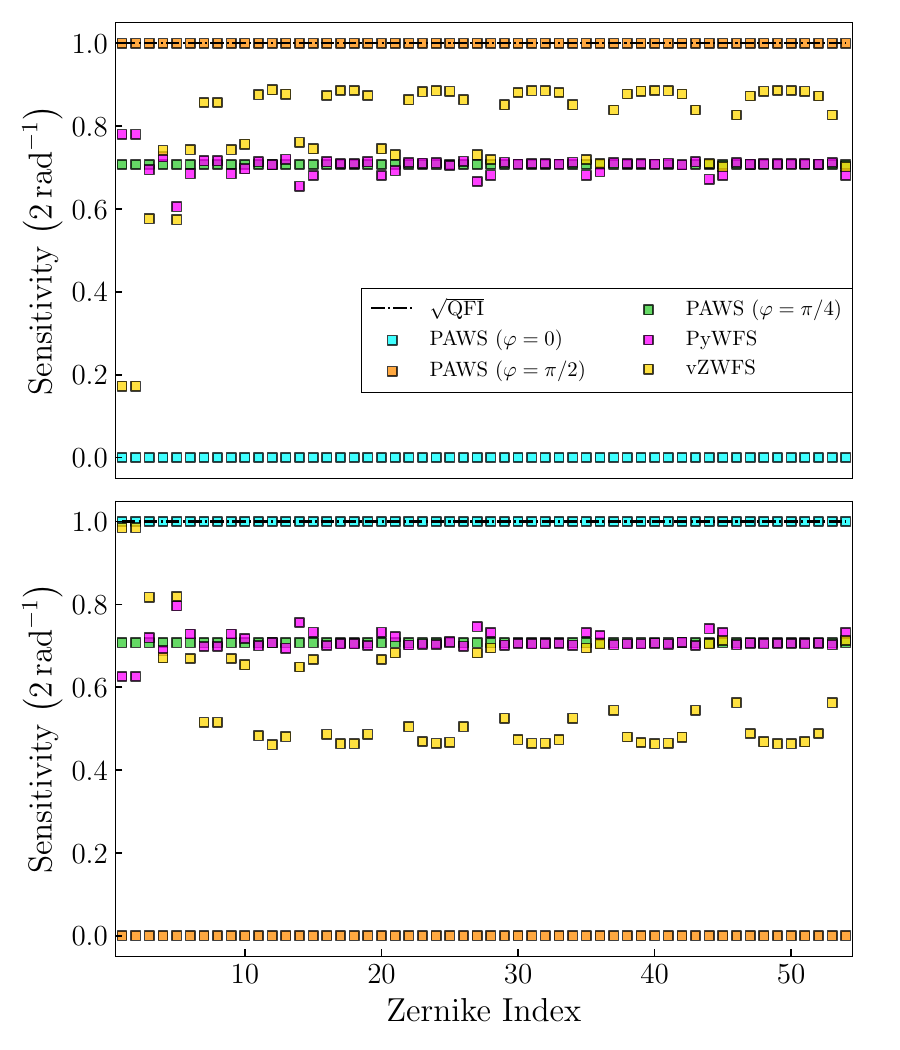}
    \caption{Photon noise-limited high-Strehl sensitivities to the first 54 non-piston (top) phase and (bottom) amplitude Zernike coefficients for various WFS. Sensitivities were calculated using the method found in \citet{chambouleyron_modeling_2023}, which is equivalent to calculating the Fisher information \citep{haffert2023}. The QFI referenced in each plot corresponds to coefficient type: (top) phase and (bottom) amplitude.}
    \label{fig:single-aperture-sensitivity}
\end{figure}

Sensitivity curves for all three sensors to the first 54 high-order Zernike polynomials are shown in Figure \ref{fig:single-aperture-sensitivity}. Our vertical axis is the square root of each sensor's Fisher information, a quantity equivalent the sensitivity metric that is often employed by the wavefront sensing community \citep{chambouleyron_modeling_2023, haffert2023}. One will notice that PyWFS performs similarly to $\varphi=\pi/4$ PAWS, suggesting that pyramid sensors are already nearly optimal instruments. While it is true that PyWFS is highly sensitive, there is no easy method to tease apart the phase and amplitude signals. In contrast, equations \eqref{eq:paws-intensities} provide a fast and straightforward methodology. Moreover, PyWFS' FIM must be more highly correlated than PAWS.

We also see that vZWFS possesses relatively poor sensitivity to log-amplitudes. This is unsurprising as that sensor induces $\pm\pi/2$ phase shifts, which are best suited for phase estimation. According to \eqref{eq:fundamental-tradeoff} this necessarily entails lost sensitivity to amplitude; we are simply allocating less light toward usable interference with amplitude perturbations. A vZWFS with $\pm\pi/4$ phase shifts should offer more equitable sensitivity.

\section{Conclusion}
\label{sec:conclusion}
We have evaluated fundamental quantum limits to simultaneous phase and amplitude estimation and presented PAWS, a readily-manufactureable WFS architecture that saturates these bounds. PAWS can tune its sensitivity to phase and amplitude aberrations via an adjustable half-wave plate and is easily achromatized using geometric phase shifters. While only optimal for circular pupils, we explained how a single-mode converter placed prior to our setup can enable quantum-limited sensing on arbitrary apertures. This approach also provides maximal sensitivity to petal modes, whose sensing and correction is critical on upcoming extremely large telescopes.

Future work will numerically investigate PAWS' robustness against electronic noises and inter-modal crosstalk. The former is a particular nuisance due to the low flux often afforded to wavefront sensors. Conversely, we expect the latter's influence to be rather tame, because it just leakage between piston and its orthogonal complement that affects performance. Evaluating wavefront reconstruction schemes such as maximum-likelihood \citep{barrett_maximum-likelihood_2007} is also planned. Finally, there is realizing PAWS experimentally, an effort that we are actively pursuing.

%
%

\bibliography{references/phase, references/telescopes, references/wfs, references/bounds, references/misc}

\begin{appendix}
\section{Holevo Cramér-Rao Bound: Proof of Inequality \eqref{eq:holevo-bound}\label{app:sec:HCRB}}
Let \begin{equation}
    \label{app:eq:cost}
    C = \begin{bmatrix}
        \mathcal{C}_1 & 0 & 0 & & 0\\
        0& \mathcal{C}_2 & 0 &\dots & 0\\
        0 & 0 & \mathcal{C}_3 & & 0\\
        & \vdots & & \ddots & 0\\
        0 & 0 & 0 & 0 & C_K
    \end{bmatrix}
\end{equation} be a non-negative and block-diagonal matrix where \begin{equation}
    C_k = \begin{bmatrix}
        \cos^2\Theta & 0\\
        0 & \sin^2\Theta
    \end{bmatrix}
\end{equation} is yet another matrix describing the relative importance (weighting) of phase and amplitude estimation with respect to the mode $\chi_k$. We also enforce $0\leq\Theta\leq\pi/2$. Above, cosine represents our weight on the phase and the sine our weight on the amplitude. The vector we seek to estimate thus interleaves the phase and amplitude coefficients, where those parameters connected to the same mode are directly adjacent. Notice also that the square root of this matrix simply removes the squares from each nonzero term.

We now introduce \begin{equation}
    \label{eq:X}
    \hat{X}=\left[\hat{X}^{\phi_{k_1}}, \hat{X}^{l_{k_1}}, \hat{X}^{\phi_{k_2}}, \hat{X}^{l_{k_2}}, \dots, \hat{X}^{\phi_{k_K}}, \hat{X}^{l_{k_K}}\right]^T,
\end{equation} which is a vector of Hermitian operators, each having dimension $(K+1)\times (K+1)$, that act on the space spanned by $\{\psi\}\cup\{\chi_k|k=k_1,k_2,\dots k_K\}$. Operators \eqref{eq:X} are related to Holevo's bound via the relation 
\begin{subequations}
\label{eq:hcrb-optimization}
    \begin{align}
    \label{app:eq:hcrb-value}
    H &= \underset{\hat{X}}{\text{min}}\left\{\text{tr}\left[C\bra{\psi}\hat{X}\hat{X}^T\ket{\psi}\right] + \left\Vert\sqrt{C}\text{Im}\bra{\psi}\hat{X}\hat{X}^T\ket{\psi}\sqrt{C}\right\Vert_1 \right\}\\
    \label{app:eq:hcrb-constraints}
    &\text{s.t }\begin{cases}
        2\text{Re}\bra{\nabla\psi}\hat{X}^T\ket{\psi}=I \\
        \bra{\psi}\hat{X}\ket{\psi} = 0
    \end{cases},
\end{align}
\end{subequations} which is a simplified version of the general optimization problem \citep{demkowicz-dobrzanski_multi-parameter_2020} that reflects our choice of weight matrix and state purity. Here ``Im" denotes an imaginary component, ``tr"  the matrix trace, $||\cdot||_1$ is the sum of absolute values of a matrix's eigenvalues, and $I$ is the $K\times K$ identity matrix.

Optimizing \eqref{app:eq:hcrb-value} may appear daunting, but it turns out that equation \eqref{app:eq:hcrb-constraints} entirely specifies the operators $\hat{X}$. To see this, express each element of $\hat{X}$ in the $\chi$-representation and denote the $(n,m)^{\text{th}}$ element of $\hat{X}^\gamma$ by $X^\gamma_{nm}$. In this basis our second constraint requires $X^\gamma_{00} = 0$ while the first constraint mandates\begin{align}
    \label{eq:X-ampl-real}
    2\text{Re}\left(X_{l_k 0}^\gamma\right) &= \delta_{a_k \gamma},\\
    \label{eq:X-ampl-imag}
    \text{Im}\left(X_{l_k 0}^\gamma\right) &= 0,\\
    \label{eq:x-phase-real}
    \text{Re}\left(X_{\phi_k 0}^\gamma\right) &= 0\\
    \label{eq:x-phase-imag}
    2\text{Im}\left(X_{\phi_k 0}^\gamma\right) &= \delta_{\alpha_k \gamma}.
\end{align} These equations annihilate most of the first column, leaving only a single term that is either $1$, for amplitude indices ($a_k$), or $i$, for phase indices ($\alpha_k$). Hence, \begin{subequations}
\begin{align}
    \left[\langle\psi,\hat{X}\hat{X}^T\psi\rangle\right]_{nm} &= \langle\psi,\hat{X}^n\hat{X}^m\psi\rangle\\
    \label{app:eq:X-elements}
    &= \frac{1}{4}\begin{cases}
    1, & n = m\\
    i, & (\exists p,q) (n=l_{p} \text{ and } m=\phi_{q})\\
    -i, & (\exists p,q)(n=\phi_{p} \text{ and } m=l_{q})
    \end{cases}.
\end{align}
\end{subequations} 

Equation \eqref{app:eq:X-elements} implies that $\langle\psi,\hat{X}\hat{X}^T\psi\rangle$ is equivalent to \begin{equation}
    \langle\psi,\hat{X}\hat{X}^T\psi\rangle = \begin{bmatrix}
        A & 0 & 0 & & 0\\
        0& A & 0 &\dots & 0\\
        0 & 0 & A & & 0\\
        & \vdots & & \ddots & 0\\
        0 & 0 & 0 & 0 & A
    \end{bmatrix},
\end{equation} where \begin{equation}
    A = \begin{bmatrix}
        1 & i \\
        -i & 1 \\
    \end{bmatrix}.
\end{equation}The minimization is thus accomplished automatically. Directly calculating the objective function finally yields \begin{equation}
    H = \frac{K}{4N}\left[1 + \sin(2\Theta)\right],
\end{equation}where the first term corresponds to the trace and the second to the trace-norm.

Note that this calculation assumed a fixed polarization, but that the bound remains the same for unpolarized light. In that case the $\hat{X}^\gamma$ operators are promoted to dimension $4K\times4K$, since we must account for both polarization states, and the matrix $\langle\psi,\hat{X}\hat{X}^T\psi\rangle$ must be replaced as \begin{equation}
    \label{app:eq:new-X}
    \langle\psi,\hat{X}\hat{X}^T\psi\rangle \rightarrow \frac{\langle\psi_{\circlearrowright},\hat{X}\hat{X}^T\psi_{\circlearrowright}\rangle + \langle\psi_{\circlearrowleft},\hat{X}\hat{X}^T\psi_{\circlearrowleft}\rangle}{2},
\end{equation} where the subscripts $\circlearrowright$ and $\circlearrowleft$ indicate right- and left-hand circular polarization, respectively. Upon, choosing a basis given by the $\chi$ modes tensored with each of the two circular polarizations, it is easy to see that the new $\hat{X}^\gamma$ operators are given by the direct sum of the single-polarization matrix with itself. This property doubles the 'tr' and $\Vert\cdot\Vert_1$ terms in our objective function, but is canceled by the $1/2$ in equation \eqref{app:eq:new-X}, thus yielding the same result.

\section{Proof that Direct Imaging in the Pupil is Optimal for Sensing Amplitude}
\label{app:sec:amplitude}
Let $p(\mathbf{x})$ denote the probability density for a photon arrival at $\mathbf{x}\in\mathcal{P}$. Direct calculation yields an FIM \begin{align}
    F_{nm} &= \int_{\mathcal{P}}\frac{\left[\partial_{l_n}p(\mathbf{x})\right]\left[\partial_{l_m}p(\mathbf{x})\right]}{p(\mathbf{x})}d\mathbf{x}\\
    &=4\int_{\mathcal{P}}\frac{\text{Re}\left[\partial_{l_n}\psi(\mathbf{x})\psi(\mathbf{x})\right]\text{Re}\left[\partial_{l_m}\psi(\mathbf{x})\psi(\mathbf{x})\right]}{|\psi(\mathbf{x})|^2}d\mathbf{x}\\
    &=4\int_{\mathcal{P}}\left[\partial_{l_n}\psi(\mathbf{x})\right]\left[\partial_{l_n}\psi(\mathbf{x})\right]d\mathbf{x}\\
    &=4\delta_{nm},
\end{align}which is exactly the QFIM.

\section{Proof that PAWS is Holevo-Optimal\label{app:sec:HCRB-PAWS}}
Let us first evaluate the information carried by each polarization state. Indeed, through a direct calculation we find an FIM\begin{align}
    \mathcal{F}_{\mu_n\nu_k}^\pm &= \sum_n\frac{\partial_{\mu_n}P_n\partial_{\nu_k}P_n}{P_n}\\
    \notag&=2\int_{\mathcal{P}}\frac{\text{Re}\left[\partial_{\mu_n}\psi(\mathbf{x})\psi(\mathbf{x})e^{\pm i\theta}\right] \text{Re}\left[\partial_{\nu_k}\psi(\mathbf{x}) \psi(\mathbf{x}) e^{\pm i\theta}\right]}{\left|\psi(\mathbf{x})e^{\pm i\theta}\right|^2 }d\mathbf{x}\\
    &=2\int_{\mathcal{P}}\text{Re}\left[\partial_{\mu_n}\psi(\mathbf{x})e^{\pm i\theta}\right] \text{Re}\left[\partial_{\nu_k}\psi(\mathbf{x})e^{\pm i\theta}\right]d\mathbf{x}\\
    &=2\text{Re}\left(i^\mu e^{\pm i\theta}\right) \text{Re}\left(i^\nu e^{\pm i\theta}\right)\int_{\mathcal{P}}e^{2l(\mathbf{x})}\chi_k(\mathbf{x})\chi_l(\mathbf{x})d\mathbf{x}\\
    &=2\text{Re}\left(i^\mu e^{\pm i\theta}\right) \text{Re}\left(i^\nu e^{\pm i\theta}\right)\delta_{nk}
\end{align}and therefore a complete FIM
\begin{align}
    \notag \mathcal{F}_{\mu_n\nu_k} &= \mathcal{F}_{\mu_n\nu_k}^+ +\mathcal{F}_{\mu_n\nu_k}^-\\
    &= 2\delta_{nk}\left[\text{Re}\left(i^\mu e^{+i\theta}\right)\text{Re}\left(i^\nu e^{+i\theta}\right)+\text{Re}\left(i^\mu e^{-i\theta}\right)\text{Re}\left(i^\nu e^{-i\theta}\right)\right].
\end{align}Note that greek letters tell us whether we differentiating with respect to a phase or log-amplitude coefficient. Let us now transition this formula to matrix form. We see that the FIM has a block representation \begin{equation}
    \label{app:eq:paws-fim}
    \mathcal{F}=\begin{bmatrix}
        D_1 & 0 & 0 & & 0\\
        0& D_2 & 0 &\dots & 0\\
        0 & 0 & D_3 & & 0\\
        & \vdots & & \ddots & 0\\
        0 & 0 & 0 & 0 & D_K
    \end{bmatrix},
\end{equation} where \begin{equation}
    D_n = 4\begin{bmatrix}
        \sin^2\theta & 0 \\
        0 & \cos^2\theta \\
    \end{bmatrix}.
\end{equation}Hence, each block matrix is identical. They are also easily inverted: \begin{equation}
    D_n^{-1} = \frac{1}{4}\begin{bmatrix}
        \frac{1}{\sin^2\theta} & 0 \\
        0 & \frac{1}{\cos^2\theta}\\
    \end{bmatrix}.
\end{equation} Equation \ref{app:eq:paws-fim} is inverted similarly. Hence, PAWS' CRB is given as \begin{equation}
    \text{tr}\left(\mathcal{C}\mathcal{F}^{-1}\right) = \frac{K}{4}\left(\frac{\cos^2\Theta}{\sin^2\theta} + \frac{\sin^2\Theta}{\cos^2\theta}\right).
\end{equation} One can now verify that choosing \begin{equation}
    \theta = \tan^{-1}\sqrt{\cot\Theta},
\end{equation} results in \begin{equation}
    \text{tr}\left(\mathcal{C}\mathcal{F}^{-1}\right) = \frac{K}{4}\left[1+\sin(2\Theta)\right],
\end{equation} which equals the fundamental limit.
\end{appendix}
\end{document}